\pdfoutput=1

\documentclass[11pt]{article}

\usepackage{EMNLP2022}

\usepackage{times}
\usepackage{latexsym}

\usepackage[T1]{fontenc}

\usepackage[utf8]{inputenc}

\usepackage{microtype}

\usepackage{inconsolata}

\usepackage{bm}
\usepackage{array}
\usepackage{xspace}
\usepackage{amssymb}
\usepackage{enumitem}
\usepackage{booktabs}
\usepackage{graphicx}
\usepackage{multirow}
\usepackage{multicol}
\usepackage{makecell}
\usepackage{subfigure}
\usepackage[normalem]{ulem}
\usepackage[ruled]{algorithm2e}

\setlist[itemize]{leftmargin=5mm, itemsep=0mm}

\useunder{\uline}{\ul}{}

\newcommand{\name}{LeSTM\xspace}
\newcommand{\ie}{\emph{i.e.,}\xspace}
\newcommand{\eg}{\emph{e.g.,}\xspace}

\newcommand{\aka}{\emph{a.k.a.,}\xspace}
\newcommand{\etal}{\emph{et al.}\xspace}
\newcommand{\paratitle}[1]{\vspace{1ex}\noindent{\bf #1}}

\title{Lexicon-Enhanced Self-Supervised Training for \\ Multilingual Dense Retrieval}

\author{
  Houxing Ren$^1$\thanks{~~Work done during internship at Microsoft STCA.} \quad Linjun Shou$^2$ \quad Jian Pei$^3$ \quad Ning Wu$^2$ \quad Ming Gong$^2$ \quad Daxin Jiang$^2$\thanks{~~Corresponding author.} \\
  $^1$School of Computer Science and Engineering, Beihang University \\
  $^2$Microsoft STC Asia \\
  $^3$Duke University, Durham, NC, USA 27705 \\
  renhouxing@buaa.edu.cn~~\{lisho,wuning,migon,djiang\}@microsoft.com~~j.pei@duke.edu
}

\begin{document}
\maketitle

\begin{abstract}
Recent multilingual pre-trained models have shown better performance in various multilingual tasks. However, these models perform poorly on multilingual retrieval tasks due to lacking multilingual training data. In this paper, we propose to mine and generate self-supervised training data based on a large-scale unlabeled corpus. We carefully design a mining method which combines the sparse and dense models to mine the relevance of unlabeled queries and passages. And we introduce a query generator to generate more queries in target languages for unlabeled passages. 
Through extensive experiments on Mr. TYDI dataset and an industrial dataset from a commercial search engine, we demonstrate that our method performs better than baselines based on various pre-trained multilingual models. Our method even achieves on-par performance with the supervised method on the latter dataset.
\end{abstract}

\section{Introduction} \label{sec:intro}

Information Retrieval~(IR) aims to retrieve relevant passages for a given query, which plays a critical role in many industry scenarios such as Open-Domain Question Answering (QA)~\cite{lee2019latent} and Web Search~\cite{bajaj2016ms}. 
Traditionally, bag-of-words~(BOW) retrieval systems such as TF-IDF and BM25~\cite{robertson2009probabilistic} were widely used, which mainly depend on keyword matching between queries and passages. 
With the development of large-scale pre-trained language models~(PLMs)~\cite{vas2017attention,devlin2018bert} such as BERT, dense retrieval methods~\cite{lee2019latent,karpukhin2020dense} show quite effective performance. 
These methods usually employed a dual-encoder architecture to encode both queries and passages into dense embeddings and then perform approximate nearest neighbor searching~\cite{johnson2019billion}.

Recently, some works found that dense retrievers perform poorly in the zero-shot multilingual settings~\cite{zhang2021mr} due to the distributional shift. To boost the performance of dense retrievers, some previous methods for cross-domain retrieval can be directly adopted to unsupervised multilingual dense retrieval. There are two important kinds: 1) generating training data in target languages. For example, Kulshreshtha \etal applied self-training to generate labeled data and further proposed back-training~\cite{kulshreshtha2021back} to obtain more high-quality data. QGen~\cite{ma2020zero} proposed to use a query generator to generate in-domain queries. 2) leveraging sparse retrievers, which is more effective in the unsupervised setting, to enhance dense retrievers. For example, SPAR~\cite{chen2021salient} proposed to distill knowledge from BM25 to the dense model and LaPraDoR~\cite{xu2022laprador} proposed to enhance the dense model by multiplying the similarity with the BM25 score. 

However, there are three major problems when directly adopting these methods to multilingual dense retrieval.
First, zero-shot multilingual query generators suffer from grammatical adjustment and accidental translation problems~\cite{xue2020mt5}. As a result, zero-shot query generators only provide little help in bridging the gap among different languages.
Second, hybrid dense and sparse models such as LaPraDoR and SPAR get high latency in the inference stage\footnote{The latency of dense retriever on GPU is 32ms and the latency of BM25 on CPU is 36ms~\cite{gao2021coil}.}.
Finally, dense retrieval is different from other tasks, it not only needs positive query-passage pairs but also needs negative query-passage pairs~\cite{xiong2020approximate}. However, previous methods such as the back-training focus on positive pairs and simply take the top passages of BM25 as negative passages.

Although training data in target languages is very expensive, unlabeled queries and passages can be easily obtained from search engines such as \emph{Google} and \emph{Bing}. 
In this paper, we propose a novel method that augments data in target languages by combining sparse and dense models, namely \name, which stands \uline{L}exicon-\uline{e}nhanced \uline{S}elf-supervised \uline{T}raining for \uline{M}ultilingual dense retrieval. First, as we mentioned above, sparse retrievers mainly depend on keyword matching between queries and passages and dense retrievers mainly depend on the language modeling ability of pre-trained models, which indicates the sparse and dense models perform retrieval in different aspects~\cite{chen2021salient}. In addition, the sparse–dense hybrid retriever is significantly better than both sparse and dense models~\cite{zhang2021mr,ma2021replication}. Both can demonstrate that sparse and dense models notice different characteristics and are complementary. Therefore, we craft a lexicon-enhanced retrieval module to mine positive and negative passages for each unlabeled query in target languages, which leverages the retrieval results of both sparse and dense models. We treat passages that both sparse and dense models regard are relevant as positive passages, and passages that one model regards are relevant but the other regards are irrelevant as negative passages. 

Furthermore, we employ a query generator to generate queries for passages in target languages due to the limited number of unlabeled queries. The query generation methods have been shown to significantly improve the performance of retrieval models in the monolingual setting~\cite{kulshreshtha2021back,ma2020zero}. Considering the grammatical adjustment and accidental translation problems, we first use the mined positive query-passage pairs to train a query generator. Then, we use the trained model to generate more queries in target languages. Considering that there may exist more relevant passages to the generated queries, we use both sparse and dense retrievers to filter the generated samples.
Finally, using only unlabeled data from target languages, \name iteratively mines query passage pairs by the lexicon-enhanced retriever and generator, trains a new better retriever and query generator using these mined pairs, mines again for better query passage pairs, and repeats.

In summary, our contributions are as follows.
\begin{itemize}
    \item To the best of our knowledge, our approach is the first attempt to combine sparse and dense retrievers to mine high-quality positive and negative query-passage pairs for the multilingual dense retriever.
    \item We propose to use a query generator to expand the unlabeled queries in target languages and an iterative training paradigm is introduced to further enhance the dense retriever and generator.
    \item Extensive experiments on two datasets show the effectiveness of our proposed approach. In particular, experiments demonstrate that our method is model-agnostic, they are effective on various pre-trained language models.
\end{itemize}

\section{Related Work}

\paratitle{Information Retrieval.} Information retrieval aims to search relevant passages from a large corpus for a given query. Traditionally, researchers use bag-of-words~(BOW) based methods such as TF-IDF and BM25~\cite{robertson2009probabilistic}. These methods use a sparse vector to represent the text, so we call them sparse retrievers. Recently, some studies use neural networks to improve the sparse retriever. For example, DocTTTTTQuery~\cite{nogueira2019doc2query} proposes to expand the document to narrow the vocabulary gap and DeepCT~\cite{dai2019context} generates a weight for each term to emphasize the import terms. 

In contrast to sparse retrievers, dense retrievers usually encode both queries and passages into dense vectors whose lengths are much less than sparse vectors. There are two kinds of dense retrieval methods: 1) pre-training with unlabeled data and 2) fine-tuning with labeled data. 
For pre-training, ORQA~\cite{lee2019latent} proposes Inverse Cloze Task~(ICT) which aims to predict the context of a given sentence, and REALM~\cite{guu2020realm} proposes to predict the masked text based on an end-to-end retriever-reader model. 
Furthermore, SEED~\cite{lu2021less}, Condenser~\cite{gao2021condenser}, and coCondenser~\cite{gao2021unsupervised} propose pre-training tasks to encode more information into the dense vectors.
For fine-tuning, one major method is how to incorporate hard negative samples during training, including static sparse hard negative samples~\cite{karpukhin2020dense,luan2020sparse} and dynamic dense hard negative samples~\cite{xiong2020approximate,zhan2021optimizing}.
Another major method is training the retriever with a cross-attention encoder jointly, including extractive reader~\cite{yang2020retriever}, generative reader~\cite{izacard2020distilling}, and cross-encoder re-ranker~\cite{qu2021rocketqa,ren2021rocketqav2,zhang2021adver}. In addition, some works trade time for performance by using multiple vectors to represent the passage~\cite{khattab2020colbert,humeau2020poly,tang2021improving,zhang2022multi}.

\paratitle{Cross-lingual~(domain) Retrieval.}
These tasks aims to investigate the retrieval capabilities under cross-lingual~\cite{zhang2021mr,asai2020xor} or cross-domain~\cite{thakur2021beir} setting.
The methods for these tasks can be divided into two main categories: model transfer methods and data transfer methods.

The model transfer methods for cross-domain focus on pre-training sentence representation. For example, GTR~\cite{ni2021large} and CPT~\cite{neelakantan2022text} propose that scaling up the model size can significantly improve the performance of dense models. Contriever~\cite{izacard2021towards} and LaPraDoR~\cite{xu2022laprador} propose to use contrastive learning to learn sentence aware representation. For cross-lingual, they focus on learning multilingual representations by pre-training~\cite{lample2019cross,chi2020infoxlm,feng2022language} such as mBERT~\cite{devlin2018bert} and XLMR~\cite{conneau2019unsupervised}.

The data transfer methods mainly focus on obtaining noisy training data in the target domain or target languages. For example, Back-training~\cite{kulshreshtha2021back} and QGen~\cite{ma2020zero} propose to use a query generator to generate in-domain queries. CORA~\cite{asai2021one} leverages a generator to help mine retrieval training data and DR.DECR~\cite{li2021learning} mines lots of parallel data to perform cross-lingual distillation. 
\section{Preliminaries} \label{sec:background}

In this section, we give a brief review of dense retrieval and then present how to apply models to multilingual dense retrieval.

\paratitle{Overview.} Given a query $q$ and a corpus $C$, the retrieval task aims to find the relevant passages for the query from a large corpus. Usually, a dense retrieval model employs two dense encoders~(\ie BERT) $E_Q(\cdot)$ and $E_P(\cdot)$. They encode queries and passages into dense embeddings, respectively. Then, the model uses a similarity function, often dot-product, to perform retrieval:
\begin{equation}
    f(q, p) = E_Q(q) \cdot E_P(p),
\end{equation}
where $f$ denotes the similarity function, $q$ and $p$ denote the query and the passage, respectively. During the inference stage, we apply the passage encoder $E_P(\cdot)$ to all the passages and index them using FAISS~\cite{johnson2019billion} which is an extremely efficient, open-source library for similarity search. Then given a query $q$, we derive its embedding by $\bm{v}_q = E_Q(q)$ and retrieve the top $k$ passages with embeddings closest to $\bm{v}_q$.

\paratitle{Training.} The training of retrieval is metric learning essentially. The goal is to narrow the distance between the query and the relevant passages~(\aka positive passages) and widen the distance between the query and the irrelevant passages~(\aka negative passages). Let $\{q_i, p^{+}_{i}, p^{-}_{i,0}, p^{-}_{i,1}, \dots, p^{-}_{i,n}\}$ be the $i$-th training sample. It consists of one query, one positive passage, and $n$ negative passages. Then we can employ a contrastive loss function, called InfoNCE~\cite{oord2018representation}, to optimize the model:
\begin{equation} 
    \mathcal{L} = - \log \frac{e^{f(q_i, p^{+}_{i})}}{e^{f(q_i, p^{+}_{i})} + \sum_{j=0}^{n} e^{f(q_i, p^{-}_{i,j})}}.
\end{equation}

In practice, we cannot use all passages in the corpus $C$ as negative passages due to the limitation of resources. Therefore, a common practice is sampling a subset from the corpus $C$ as negative samples, and many studies focus on which distribution the negative passages sampled from is better~\cite{xiong2020approximate,qu2021rocketqa}.

\paratitle{Multilingual Setting.} This setting aims to transfer knowledge from the source language to the target languages. In this setting, only labeled data from the source language is available. And the trained model will be directly evaluated on target languages. Note that the setting is different to \emph{cross-lingual retrieval} whose queries and passages are in different languages. In this setting, the queries and the passages are in the same language and just training data from the source language~(\eg English) is available.

\section{Methodology}

\begin{figure}[t]
    \centering
    \includegraphics[width=\columnwidth]{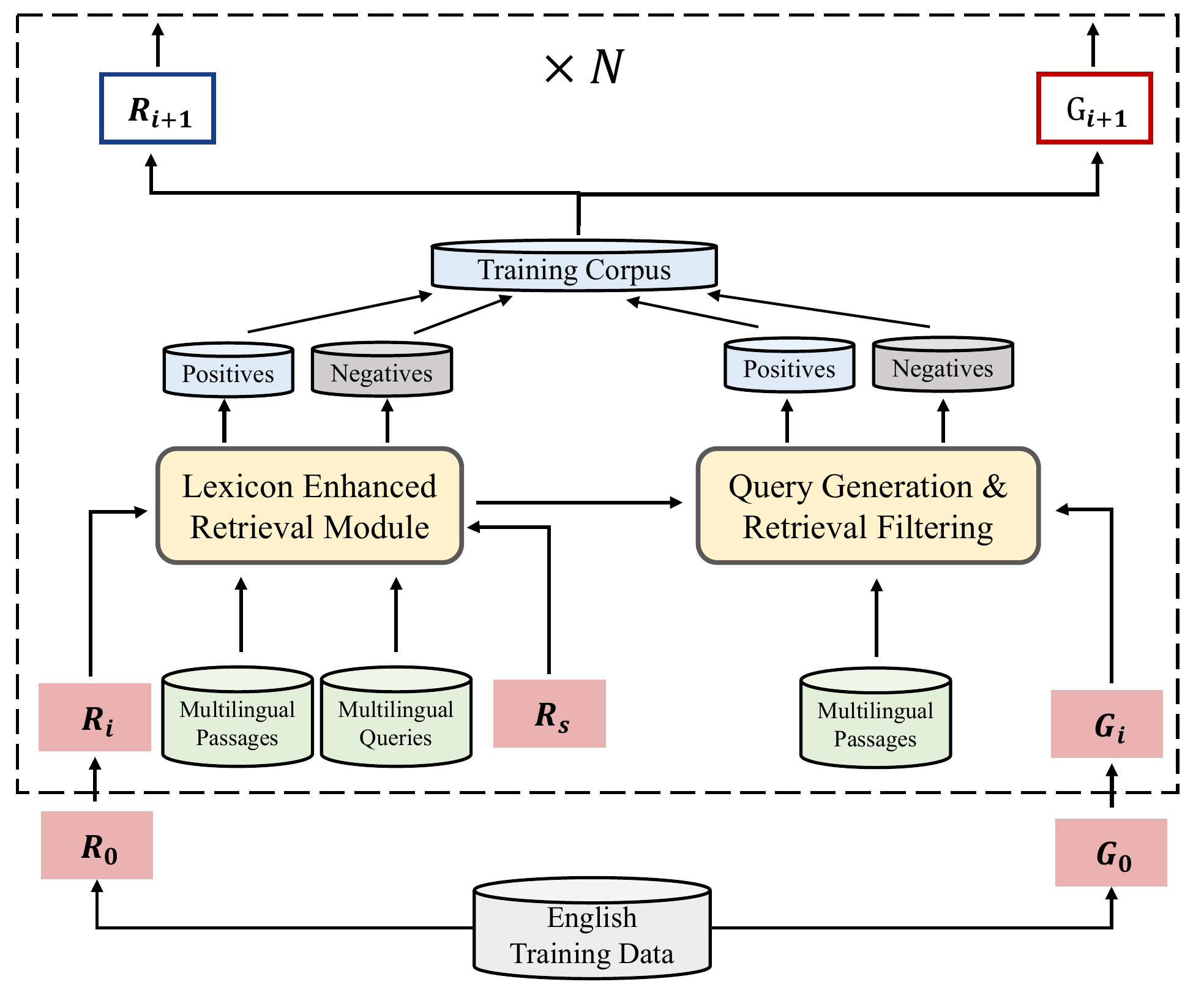}
    \caption{Overview of the training process.}
    \label{fig:overview}
\end{figure}

In this section, we present the proposed \name. The overview is presented in Figure~\ref{fig:overview}. We first present the augmentation method which combines sparse and dense retrievers. Then, we present how to use the mined data to train the query generator, generate new data, filter the generated samples, and fine-tune the dense retriever. Finally, we summarize the full training process.

\subsection{Lexicon-enhanced Retrieval Module}

In target languages, we do not have the labeled training data, but we have unlabeled queries and passages. To effectively utilize the unlabeled queries and passages, we design a mining method shown in Algorithm~\ref{alg:pm}.

This augmentation is based on the intuition that the sparse retriever and the dense retriever solve different problems and they are complementary. Specifically, the sparse retriever depends on word match, it is more effective than the dense retriever for words that do not appear in the training set. On the contrary, the dense retriever depends on neural networks, it is more effective than the sparse retriever for synonyms and semantics of the sentence. 
As a result, for a passage, if both of them regard it are relevant to the query, we then regard the passage as a positive passage. If one of them regards it as relevant but the other regards it as irrelevant, we regard the passage does not meet all conditions~(\ie keyword match and semantic match), and the passage is a hard case. 
Although we cannot judge whether it is a negative passage, we think its relevance is weaker than the positive passage. As a result, we hope the score of the positive passage is higher than the hard case and then we regard the hard case as a hard negative passage.

In practice, because the score distributions of sparse retrievers and dense retrievers are different, we use the ranking position to measure the relevance between passages and queries. Then, we present our method as follows:
\begin{enumerate}[label=(\arabic*)]
    \item We introduce two parameters to define relevant and irrelevant passages: $S$ and $L$, \ie for a query, the retriever retrieves passages and ranks them with scores, if the ranking position of a passage is less than $S$, we regard the passage is relevant to the query and if the ranking position is greater than $L$, we regard the passage is irrelevant to the query.
    \item We retrieve $L$ and $S$ passages by both sparse and dense retrievers, respectively. We define the top-$L$ passage set as $\mathbb{L}$ and the top-$S$ passage set as $\mathbb{S}$, and use subscript $s$ and $d$ to denote sparse and dense retrievers, respectively. 
    \item We traverse all passages in the corpus. For each passage, if it exists in both $\mathbb{S}_s$ and $\mathbb{S}_d$, we add it to the positive passage set; if it exists in one $\mathbb{S}$ but does not exist in another $\mathbb{L}$~(\aka exists in $\mathbb{S}_s$ but not exists in $\mathbb{L}_d$ or exists in $\mathbb{S}_d$ but does not exist in $\mathbb{L}_s$), we add it to the negative passage set. 
    \item For each mined sample, we add random negative passages like DPR~\cite{karpukhin2020dense}: 1) random passages from the corpus; 2) positive passages of other queries~(\aka in-batch negative). And we use our mined negative passage set to replace ``top passages returned by BM25'' in DPR.
\end{enumerate}

\begin{algorithm}[t] \small
    \caption{Lexicon-enhanced Retrieval Module.}
    \label{alg:pm} 
    \LinesNumbered
    \KwIn{One query $q$ and candidate passages $P$.}
    \KwOut{Positive passages and Negative passages.}
    
    Set $L$ and $S$\;
    $\mathbb{L}_s ~\&~ \mathbb{S}_s \leftarrow$ Top-L and Top-S of sparse retriever\;
    $\mathbb{L}_d ~\&~ \mathbb{S}_d \leftarrow$ Top-L and Top-S of dense retriever\;
    $P^{+} \leftarrow \varnothing$; $P^{-} \leftarrow \varnothing$\;
    \For{$p \in P$}{
        \If{$p \in \mathbb{S}_s ~\&~ p \in \mathbb{S}_d$}{
            $P^{+} \leftarrow P^{+} \cup \{p\}$
        }
        \If{$p \in \mathbb{S}_s ~\&~ p \notin \mathbb{L}_d$}{
            $P^{-} \leftarrow P^{-} \cup \{p\}$
        }
        \If{$p \notin \mathbb{L}_s ~\&~ p \in \mathbb{S}_d$}{
            $P^{-} \leftarrow P^{-} \cup \{p\}$
        }
    }
    \textbf{Return} $P^{+}$, $P^{-}$.
\end{algorithm}

To sum up, our mined training data includes a query, mined positive and negative passages, and random negative passages.

\subsection{Query Generation Module}

Due to the limited number of unlabeled queries, we leverage a query generator to generate more queries for unlabeled passages in target languages. Note that the generated queries are in the same languages as the corresponding passage.

Specifically, for a trained generator, we randomly select some passages and leverage the fine-tuned query generator to generate queries for these passages.
To tackle the noisy label problem introduced by the generator, we use both sparse and dense retrievers to filter the generated pairs. We retrieve the top-1 passage for each generated query with both sparse and dense retrievers and only accept pairs where the best passages from both sparse and dense retrievers are the corresponding passage. 
Finally, for each filtered sample, we select negative passages like DPR~\cite{karpukhin2020dense}: 
1) random passages from the corpus; 
2) top passages returned by sparse and dense retrievers~(the passages returned by dense retriever are more effective~\cite{xiong2020approximate,qu2021rocketqa});
3) positive passages of other queries.

To sum up, our generated training data includes one positive passage, generated query, random negative passages, and top passages returned by retrievers as hard negative passages.

\subsection{Model Training}

\begin{algorithm}[t] 
    \caption{The training algorithm.}
    \label{alg:overview} 
	\LinesNumbered
    \KwIn{Labeled English training data and unlabeled queries and passages in target languages.}
    Construct index for sparse retriever\;
    Initialize the dense retriever and generator with pre-trained models\;
    Train the retriever and generator with English data\;
    Build ANN index for the retriever\;
    \While{models has not converged}{
        Generating training data $\mathbb{D}_p$ with passage mining module\;
        Generating training data $\mathbb{D}_g$ with query generation module\;
        Fine-tune the generator and retriever with both $\mathbb{D}_p$ and $\mathbb{D}_g$\;
        Refresh ANN index for the retriever.
    }
\end{algorithm}

Previously, we introduced the lexicon-enhanced retrieval module and the query generation module. In this part, we present the full training process.

As shown in Algorithm~\ref{alg:overview}, 
firstly, we train the warm-up dense retriever and query generator with data in the source language. We note that the input to the generator is the positive passage, and the label is the query. 
Secondly, we generate training data in target languages with the proposed two modules.
Finally, we fine-tune the retriever and the generator with the generated data. 
Based on these steps, we can conduct iteratively generating and training procedures to improve the performance. 
Note that due to the grammatical adjustment and accidental translation problems in the zero-shot multilingual generator, we only use the lexicon-enhanced retrieval module to generate data in the first iteration.

Considering the query generator is more sensitive to the quality of data, we set $S=1$ when generating data for the query generator.

\section{Experiments}

In this section, we construct experiments to demonstrate the effectiveness of the proposed method.

\subsection{Experimental Setup}

\subsubsection{Dataset} 

\paratitle{Mr. TYDI.} The Mr. TYDI dataset~\cite{zhang2021mr} is constructed from TYDI~\cite{clark2020tydi} dataset and can be viewed as the ``open-retrieval'' condition of the TYDI dataset. It is a multilingual dataset for monolingual retrieval in 11 languages. The detailed statistics of the Mr. TYDI dataset are presented in Appendix~\ref{sec:dataset}.

\paratitle{DeepQA.} An Q\&A task dataset from one commercial Q\&A system, with 18,000 labeled cases in three languages: English~(En), German~(De), French~(Fr). Each case consists of two parts, \ie query and passage. The detailed statistics of the DeepQA dataset are presented in Appendix~\ref{sec:dataset}. 

\subsubsection{Evaluation Metrics.} 
Following Mr. TYDI, we use MRR@100 and Recall@100 as evaluation metrics, where MRR denotes the mean of reciprocal rank across queries and Recall@k denotes the proportion of queries to which the top k retrieved
passages contain positives. For DeepQA, due to the smaller size of the corpus~(only 1,220,030 passages in the corpus, for comparison, the Mr. TYDI data has 58,043,326 passages, which is times that of DeepQA), we use MRR@10 and Recall@10 as metrics.

\subsubsection{Implementation Details.} 

For the warm-up training stage, although Mr. TYDI proposed to use NQ~\cite{kwiatkowski2019natural} as English training data, we follow Xinyu~\etal~\cite{zhang2022towards} to use MS-MARCO as English training data. Xinyu~\etal find that MS-MARCO is better than NQ for zero-shot multilingual retrieval. We have further constructed experiments on NQ in Appendix~\ref{sec:nq}.

For the iteratively training stage, both the retriever and the generator are scheduled to train with 500 mini-batches in each iteration. The document index is refreshed after each iteration of training. The hyper-parameters are shown in Appendix~\ref{sec:param}.

All the experiments run on 8 NVIDIA Tesla A100 GPUs. The implementation code
is based on HuggingFace Transformers~\cite{wolf2020transformers}. For sentence embedding, we use the corresponding hidden state of the \emph{[CLS] token} for mBERT~\cite{devlin2018bert} and the average hidden states of all tokens for XLM-R~\cite{conneau2019unsupervised}. For the generator, we leverage mBART~\cite{liu2020mbart} as the pre-trained model.

\begin{table}[t] \small
\centering
\setlength\tabcolsep{4pt}
\caption{Results on Mr. TYDI test set. The best results except supervised training are in bold. We copy the results of BM25, tuned BM25, and zero-shot mBERT from \cite{zhang2022towards} and re-implement the zero-shot mBERT. $\ast$ denotes that our method significantly outperforms self-training at the level of 0.01. $\dag$ denotes that our method significantly outperforms back-training at the level of 0.01.}
\begin{tabular}{c|c|cc} \toprule
\multicolumn{2}{c|}{Methods} & MRR@100 & Recall@100 \\
\midrule \multirow{2}{*}{\shortstack{Sparse \\ Method}}
& BM25          & 32.1          & 73.2          \\ 
& (tuned)       & 33.3          & 75.8          \\ \midrule 

\multirow{7}{*}{mBERT}
& Zero-Shot     & 34.4          & 73.4          \\
& (reimpl)      & 36.5          & 73.3          \\ \cmidrule(lr){2-4}
& Self-Training & 37.2          & 78.5          \\
& Back-Training & 41.1          & 82.0          \\
& \name         & \textbf{49.0}$^{\ast \dag}$ & \textbf{83.6}$^{\ast \dag}$ \\ \cmidrule(lr){2-4}
& Supervised    & 54.6          & 87.0          \\ \midrule 

\multirow{6}{*}{XLM-R}
& Zero-Shot     & 30.4          & 74.3          \\  \cmidrule(lr){2-4}
& Self-Training & 35.0          & 78.6          \\
& Back-Training & 29.6          & 77.5          \\
& \name         & \textbf{47.2}$^{\ast \dag}$ & \textbf{82.7}$^{\ast \dag}$ \\ \cmidrule(lr){2-4}
& Supervised    & 54.5          & 87.2          \\ \midrule 

\end{tabular}

\label{tab:mrtydi}
\end{table}

\subsection{Results}

\subsubsection{Baselines}
As we investigate retrieval in the multilingual setting, in this paper, the main baselines methods include BM25, and multilingual DPR with mBERT~\cite{devlin2018bert}, XLM-R~\cite{conneau2019unsupervised} as the multilingual pre-trained model. 
Furthermore, we compare our method with two state-of-the-art domain adaption methods: self-training~\cite{yarowsky1995unsupervised} and back-training~\cite{kulshreshtha2021back}. Following Back-training,  we train the models 3 iterations with 5 epochs per iteration. Then we present the results with the best MRR@100.
In addition, we present the supervised performance as an upper limit reference. When constructing the supervised training data, we follow DPR~\cite{karpukhin2020dense} to select three kinds of negative passages.

\subsubsection{Mr. TYDI}

Table~\ref{tab:mrtydi} shows the result on Mr. TYDI. The first group is the sparse retriever, \ie BM25~\cite{robertson2009probabilistic} and tuned BM25. For each pre-trained model, the first group is the multilingual pre-trained models which are only fine-tuned on MS-MARCO data. The second block is the multilingual pre-trained models which are fine-tuned on MS-MARCO data and data augmentation method. We conduct pair t-test~\cite{hsu2014paired} between our method and other data augmentation method~(self-training and back-training). The final block is the multilingual pre-trained models which are fine-tuned on Mr. TYDI dataset. Due to the limited space, we only present the average performance among all languages in Table~\ref{tab:mrtydi} and present results for each language in Appendix~\ref{sec:lang}. 

Based on the results, we have the following findings.
Firstly, comparing the performance of domain adaption methods (the second block for each pre-trained model) and zero-shot performance, we can find that all domain adaption methods are effective. 
Secondly, comparing the three domain adaption methods, we can find that our method is better than the other methods. 
Finally, comparing our method and supervised dense retriever, we can find that the performance of our method is closed to the supervised performance on Recall@100, but is still worse than supervised performance with a clear edge on MRR@100. This indicates that the augmented data are noisy, for example, the mined passages are relevant to the queries but are not the best passages, and there may be more relevant passages for the queries. So, it is more helpful to Recall@100 but less helpful to MRR@100.

\begin{table}[t] \small
\centering
\setlength\tabcolsep{4pt}
\caption{Results on DeepQA test set. The best results except supervised training are in bold. $\ast$ denotes that our method significantly outperforms self-training at the level of 0.01. $\dag$ denotes that our method significantly outperforms back-training at the level of 0.01.}
\subtable[MRR@10]{\begin{tabular}{c|ccc|c} \toprule
Methods       & En             & De             & Fr             & Avg            \\ \midrule 
BM25          & 22.5           & 31.4           & 40.1           & 31.3           \\ \midrule 
Zero-Shot     & 24.0           & 29.4           & 37.7           & 30.3           \\ \midrule
Self-Training & 25.3           & 31.4           & 42.3           & 33.0           \\ 
Back-Training & 25.8           & 32.0           & 42.0           & 33.3           \\ 
\name          & \textbf{27.2}$^{\ast \dag}$  & \textbf{34.6}$^{\ast \dag}$  & \textbf{43.0}$^{\ast \dag}$  & \textbf{35.0}$^{\ast \dag}$  \\ \midrule
Supervised    & 23.0           & 33.9           & 39.7           & 32.2           \\ \bottomrule 
\end{tabular}}

\subtable[Recall@10]{\begin{tabular}{c|ccc|c} \toprule
Methods       & En             & De             & Fr             & Avg            \\ \midrule 
BM25          & 37.2           & 52.1           & 56.8           & 48.7           \\ \midrule 
Zero-Shot     & 39.1           & 49.3           & 56.7           & 48.4           \\ \midrule
Self-Training & 41.8           & 55.9           & 60.9           & 52.7           \\
Back-Training & 42.6           & 55.8           & 61.4           & 53.2           \\
\name          & \textbf{44.2}$^{\ast \dag}$  & \textbf{57.9}$^{\ast \dag}$  & \textbf{62.9}$^{\ast \dag}$  & \textbf{55.0}$^{\ast \dag}$  \\ \midrule
Supervised    & 38.8           & 62.8           & 63.0           & 54.7           \\ \bottomrule 
\end{tabular}}

\label{tab:deep}
\end{table}

\begin{table}[t] \small
\centering
\setlength\tabcolsep{4pt}
\caption{Performance of zero-shot dense retriever on DeepQA training set and test set.}
\begin{tabular}{l|cc} \toprule
~            & MRR@10         & Recall@10 \\ \midrule 
Training set & 35.1           & 48.5      \\ 
Test set     & 30.3           & 48.4      \\ \bottomrule 
\end{tabular}
\label{tab:train-test}
\end{table}

\subsubsection{DeepQA}

Due to the limited space, we only construct experiments on DeepQA based on mBERT. Table~\ref{tab:deep} presents the performance of all methods. As we can see, our method achieves the best performance among all the compared methods. It indicates that our method is effective for unsupervised multilingual dense retrieval.

In addition, we find that the unsupervised methods~(\ie self-training and back-training) perform better than the supervised training on MRR@10 but worse on Recall@10. A possible reason is that the size of DeepQA is small and there is a large gap between the distributions of the training data and test data. To demonstrate that, we evaluate the performance of the zero-shot dense retriever on both training data and test data. As shown in Table~\ref{tab:train-test}, there is a large gap between the MRR@10 on the training set and the test set of DeepQA. That indicates the gap between the training set of the test set is large. The dense model trained on the training set may seriously suffer from the overfitting problem. These results also indicate that our method is even more effective than supervised training when the training data in target languages is limited.

\begin{table}[t] \small
\centering
\setlength\tabcolsep{4pt}
\caption{Ablation results based on mBERT. ``LR'' denotes the lexicon-enhanced retrieval module. ``QG'' denotes the query generation module.}
\begin{tabular}{l|cc} \toprule
Methods      & MRR@100       & Recall@100    \\ \midrule
\name         & \textbf{49.0} & \textbf{83.6} \\ \midrule
w/o LR       & 46.9          & 81.6          \\
w/o LR$_+$   & 37.9          & 77.9          \\
w/o QG       & 48.1          & 83.2          \\ 
w/o ALL      & 36.5          & 73.3          \\ \bottomrule
\end{tabular}
\label{tab:ablation}
\end{table}

\subsection{Ablation Study}

In our method, we have incorporated two data augmentation modules, namely lexicon-enhanced retrieval, and query generation. Here, we would like to check how each module contributes to the final performance. We construct the ablation experiments on the Mr. TYDI data. We prepare four variants of our method that try all combinations: 
\begin{itemize}
    \item {\ul w/o LR} denotes that the retriever does not be fine-tuned with data from the lexicon-enhanced retrieval module. But the generator also is fine-tuned with data from the lexicon-enhanced retrieval module.
    \item {\ul w/o LR$_+$} denotes that both the retriever and the generator do not be fine-tuned with data from the lexicon-enhanced retrieval module. 
    \item {\ul w/o QG} denotes that the retriever does not be fine-tuned with data from the query generation module.
    \item {\ul w/o ALL} denotes without both the two modules, \aka zero-shot multilingual retrieval.
\end{itemize}

Table~\ref{tab:ablation} presents all comparison results of the four variants. Due to the limited space, we present results for each language in Appendix~\ref{sec:lang}. 
As we can see, the performance rank can be given as follows w/o ALL < w/o QG  < \name. These results indicate that both the two augmentation modules are essential to improve performance. And we can find that the lexicon-enhanced retrieval module is more effective than the query generation module, because of w/o LR < w/o QG. In addition, we find that w/o LR > w/o LR$_+$, it denotes the zero-shot multilingual query generation suffers from lots of problems and it also can demonstrate the effectiveness of the lexicon-enhanced retrieval module.

\begin{table}[t] \small
\centering
\setlength\tabcolsep{4pt}
\caption{Effect of lexicon-enhanced retrieval module.}
\begin{tabular}{l|cc} \toprule
    Methods                 & MRR@100       & Recall@100    \\ \midrule
    BM25                    & 32.1          & 73.2          \\
    Zero-shot mBERT         & 36.5          & 73.3          \\ \midrule
    \name w/o QG            & \textbf{47.5} & \textbf{82.5} \\ \midrule
    Sparse $+$ Dense        & 46.6          & 81.1          \\
    Sparse $\times$ Dense   & 40.8          & 80.4          \\
    Double Dense Retrievers & 44.5          & 81.9          \\ \midrule
    w/o HN                  & 42.5          & 80.4          \\
    w/ Sparse HN            & 43.3          & 79.2          \\ \bottomrule
\end{tabular}
\label{tab:passage}
\end{table}

\subsection{Method Analysis}

\subsubsection{Effect of Lexicon-enhanced Retrieval} 
In our lexicon-enhanced retrieval module, we combine the results of the sparse and dense retrievers to mine new training data. To show the effectiveness of our mining method, we construct the five variants~(for more conciseness, we use mBERT {\ul w/o QG + iterative refinement} as the base model):

\begin{itemize}
    \item {\ul Sparse $+$ Dense} combines results of sparse and dense retrievers by adding their scores.
    \item {\ul Sparse $\times$ Dense} combines results of sparse and dense retrievers by multiplying their scores. 
    \item {\ul Double Dense Retrievers} mines positive and negative passages with results from two dense retrievers which are trained on different data~(MS-MARCO and NQ).
    \item {\ul w/o Hard Negatives~(HN)} fine-tunes the model with mined positive passages and only in-batch negative passages.
    \item {\ul w/ Sparse Hard Negatives~(HN)} fine-tunes the model with mined positive passages, in-batch negative passages, and top passages returned by sparse retriever as negative passages.
\end{itemize}

Table~\ref{tab:passage} presents all comparison results of the five variants.  Based on the results, we have the following findings. 
Firstly, our mining method is more effective than the hybrid results of sparse and dense models. It demonstrates that our method can effectively leverage the knowledge of both sparse and dense retrievers. 
Secondly, mining data with sparse and dense retrievers are more effective than two dense retrievers. It demonstrates that sparse and dense retrievers have noticed different characteristics of retrieval.
Finally, mined negatives are more effective than sparse negatives. It demonstrates that negatives are important in dense retrieval tasks and our methods can provide more effective negatives. 

\begin{figure}[t]
    \centering
    \subfigure[$S$ in Algorithm~\ref{alg:pm}.]{
        \includegraphics[width=0.75\columnwidth]{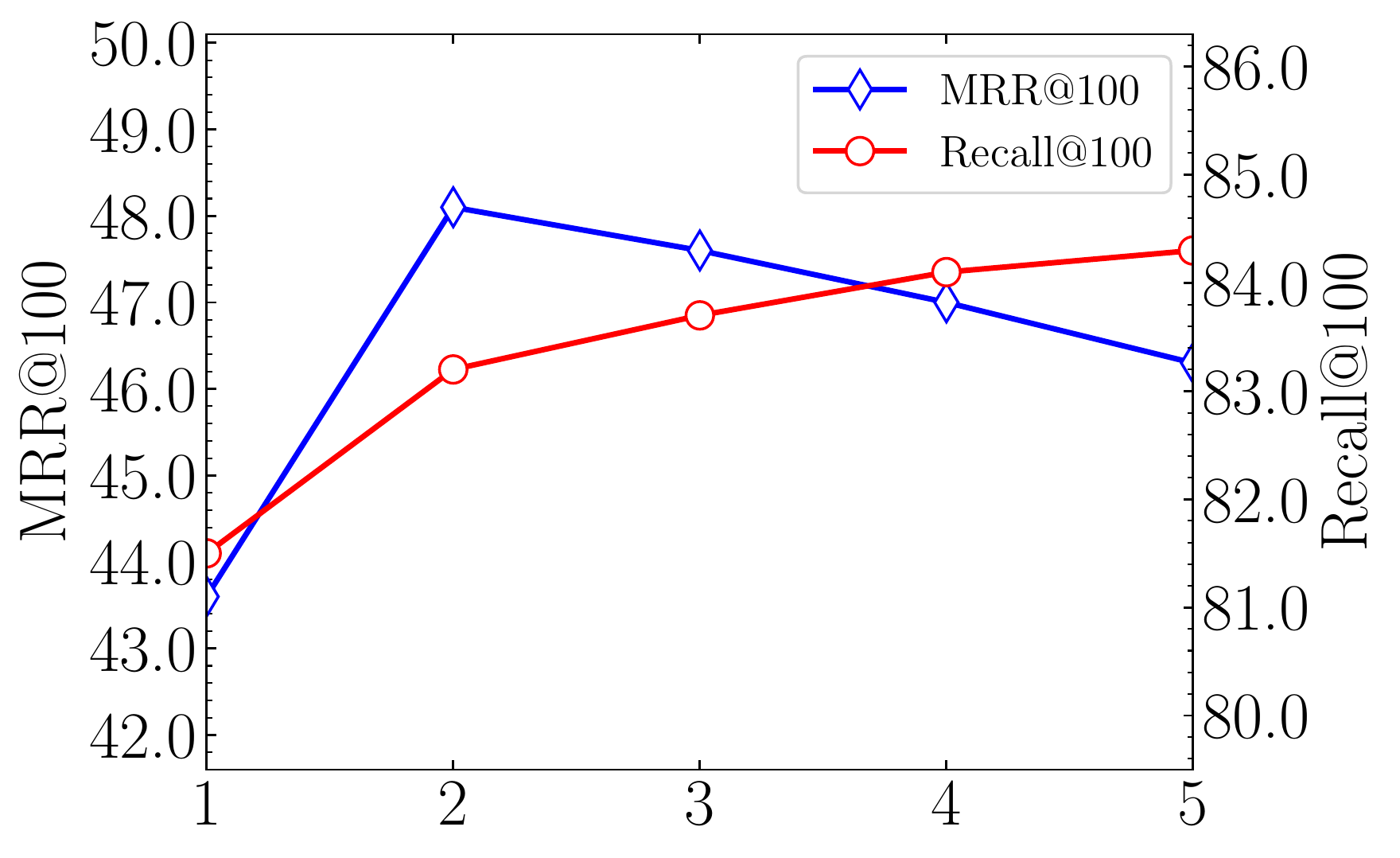}
        \label{fig:sens-thresh}
    }
    \subfigure[The number of generated queries.]{
        \includegraphics[width=0.75\columnwidth]{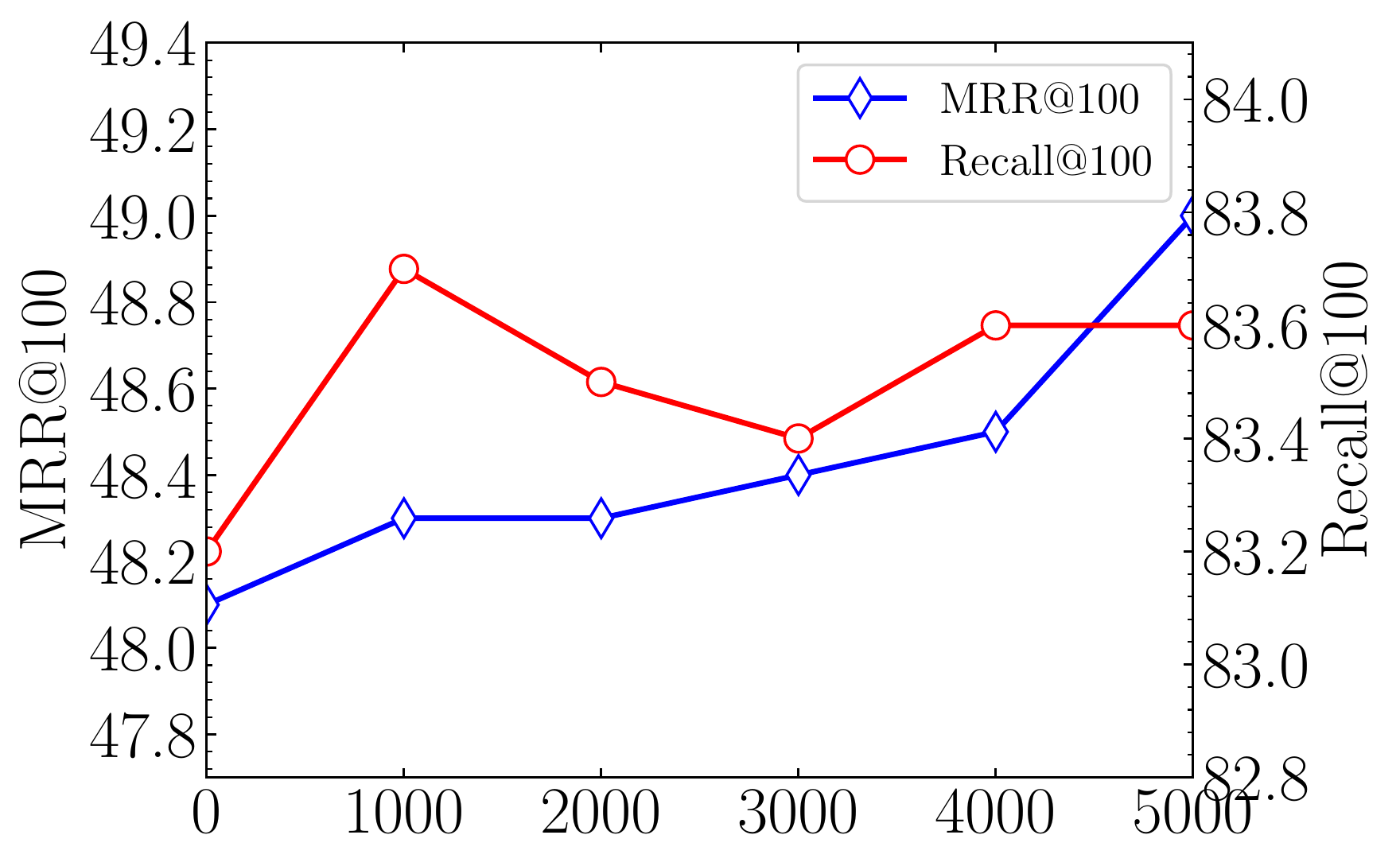}
        \label{fig:sens-gen}
    }
    \caption{Parameter sensitivity.}
    \label{fig:sensitivity}
\end{figure}

\subsubsection{Effect of Parameters} \label{sec:params}
In our method, we introduce two parameters in the lexicon-enhanced retrieval module to define relevant and irrelevant passages: $S$ and $L$. And the number of generated queries also influences the final performance. 
Here, we tune the  $S$ and $L$ based on mBERT w/o QG. We vary $S$ in the set $\{1, 2, 3, 4, 5\}$. And for more conciseness, we set $L = S \times 10$. 
In addition, we tune the number of generated queries based on mBERT. We vary the number of generated queries per language in the set $\{1000, 2000, 3000, 4000, 5000\}$.

Figure~\ref{fig:sens-thresh} presents the effect of the parameter $S$. We can observe that $S = 1$ leads to poor performance on both MRR@100 and Recall@100. Because the method with little $S$ mines few samples and leads to the overfitting problem. When we set $S > 2$, it leads to better Recall@100 but poorer MRR@100. A possible reason is that large $S$ leads to more noisy samples. As we mentioned above, noisy samples are helpful to Recall@100 but harmful to MRR@100. 

Figure~\ref{fig:sens-gen} presents the effect of the number of generated queries. As we can see, the large number of generated queries improves the MRR@100 but cannot improve the Recall@100. A possible reason is that the generated queries mainly focus on a few kinds~(\eg when or where something happened). They are helpful to MRR@100 for these kinds of queries but less helpful to both Recall@100 and MRR@100 for other kinds of queries.
\section{Conclusion}

In this paper, we propose a novel augmentation method that combines sparse and dense retrievers for multilingual retrieval.
We firstly designed a passage mining method based on the results of both sparse and dense retrievers.
After that, we utilized the mined data to train a query generation model and generate more training data.
Extensive experimental results show that the proposed method outperforms the baselines, and can significantly improve the state-of-the-art performance. 
Currently, we directly utilize a large number of unlabeled queries in target languages. As future work, we will investigate how to augment training data without any unlabeled queries in target languages.

\section{Limitations}

The limitations are summarized as follows.

\begin{itemize}
    \item The method needs unlabeled queries. For seriously rare languages, there are no unlabeled queries in search engines and we cannot perform our passage mining method in this condition. Although our query generation module can alleviate this problem, the zero-shot query generator suffers from grammatical adjustment and accidental translation problems and can only provide limited help.
    \item The method performs inconsistently on the two metrics~(MRR@100 and Recall@100). Due to the quality of augmented data, we need to set some threshold to filter the augmented data, where different parameters lead to optimal performance on different metrics.
    \item The sparse retriever is fixed during training. The fixed sparse retriever leads to the rapid convergence of the dense retriever. We believe that if both sparse and dense retrievers can be improved in the iterative process, the dense retriever may achieve better performance.
\end{itemize}

\section*{Acknowledgments}

Jian Pei’s research is supported in part by the NSERC Discovery Grant program. All opinions, findings, conclusions and recommendations in this paper are those of the authors and do not necessarily reflect the views of the funding agencies.

\bibliography{references}
\bibliographystyle{acl_natbib}

\clearpage

\section*{Appendix}
\appendix

\section{Dataset Statistics} \label{sec:dataset}

\paratitle{Mr.TYDI.} Mr. TYDI is a multilingual retrieval benchmark based on the TYDI dataset. Mr. TYDI covers 11 languages. The corpus for each language is drawn from Wikipedia, and the query and judgments are prepared by native speakers of that language. Table~\ref{tab:stat} presents statistics of the Mr. TYDI dataset, copied from the original paper.

\paratitle{DeepQA.} DeepQA is a Q\&A dataset from one commercial Q\&A system, with 18,000 labeled cases in three languages. Each case consists of two parts, \ie query and passage. The following briefly describes how the data is collected. Firstly, for each query, the top 10 relevant documents returned by the search engine are selected to form <query, url> pairs; Then passages are further extracted from these documents to form <query, url, passage> triples; These <query, passage> pairs are sampled and sent to crowd sourcing judges. Specifically, each <query, passage> pair is required to get judged by three judges. Those cases with more than 2/3 positive labels will get positive labels, otherwise negative. The detailed statistics of the DeepQA dataset are presented in Table~\ref{tab:deep-stat}. 

\begin{table}[h] \small
\centering
\caption{Statistics for DeepQA: number of queries (\# Q), judgments (\# J), and the number of passages.}
\begin{tabular}{l|rr|rr|r} \toprule
    \multirow{2.5}{*}{L} & \multicolumn{2}{c|}{Train} & \multicolumn{2}{c|}{Test} & \multirow{2.5}{*}{Corpus Size} \\ \cmidrule(lr){2-5}
    ~ & \# Q & \# J & \# Q & \# J & ~ \\ \midrule
    En & 4,437 & 6,022 & 1,703 & 1,978 & 741,840 \\
    De & 2,401 & 3,977 & 910 & 1,023 & 385,440 \\
    Fr & 2,976 & 4,000 & 936 & 1,000 & 92,750 \\ \midrule
    All & 9,814 & 13,999 & 3,549 & 4,001 & 1,220,030 \\ \bottomrule
\end{tabular}
\label{tab:deep-stat}
\end{table}

\begin{table}[t] \small
\centering
\caption{Hyper-parameters.}
\begin{tabular}{c|l|c} \toprule
     ~ & Parameters & Value \\ \midrule
     ~ & Max Query Length & 32 \\
     ~ & Max Passage Length & 128 \\ \midrule
     \multirow{7}{*}{\shortstack{Train \\ warm-up \\ retriever}} & Learning Rate & 1e-5 \\
     ~ & Batch Size & 128 \\
     ~ & Negative Size & 255 \\
     ~ & Optimizer & AdamW \\ 
     ~ & Scheduler & Linear \\
     ~ & Warmup Proportion & 0.1 \\
     ~ & Training Epoch & 3 \\ \midrule
     \multirow{6}{*}{\shortstack{Train \\ warm-up \\ generator}} & Learning Rate & 1e-5 \\
     ~ & Batch Size & 64 \\
     ~ & Optimizer & AdamW \\ 
     ~ & Scheduler & Linear \\
     ~ & Warmup Proportion & 0.1 \\
     ~ & Training Epoch & 1 \\ \midrule
     \multirow{10}{*}{\shortstack{Iteraively \\ training of \\ retriever}} & Learning Rate & 1e-6 \\
     ~ & Batch Size & 128 \\
     ~ & Negative Size & 255 \\
     ~ & Optimizer & AdamW \\ 
     ~ & Scheduler & Linear \\
     ~ & Warmup Proportion & 0.1 \\
     ~ & Training Epoch & 5 \\
     ~ & $S$ in Algorithm~\ref{alg:pm} & 2 \\
     ~ & $L$ in Algorithm~\ref{alg:pm} & 20 \\
     ~ & \# of Generated Queries & 5000 \\
     ~ & \# of Iteration & 3 \\\midrule
     \multirow{6}{*}{\shortstack{Iteraively \\ training of \\ generator}} & Learning Rate & 1e-5 \\
     ~ & Batch Size & 64 \\
     ~ & Optimizer & AdamW \\ 
     ~ & Scheduler & Linear \\
     ~ & Warmup Proportion & 0.1 \\
     ~ & Training Epoch & 5 \\
     ~ & \# of Iteration & 3 \\\bottomrule
\end{tabular}
\label{tab:params}
\end{table}

\begin{table}[!ht] \small
\centering
\caption{Efficiency Report.}
\begin{tabular}{l|l|c} \toprule
     \multirow{5}{*}{Training} & Warm-up & 0.5h \\
     ~ & Per Iteration & 0.2h \\
     ~ & Index Refresh & 1.7h \\
     ~ & Generate Queries & 0.5h \\ 
     ~ & Overall & 8.9h \\ \midrule
     \multirow{3}{*}{Inference} & Build Index & 1.7h \\
     ~ & Query Encoding & 40ns \\
     ~ & Dense Retrieval & 2ms \\ \midrule
\end{tabular}
\label{tab:efficiency}
\end{table}

\section{Hyper-parameters} \label{sec:param}

We have analyzed the parameters of our method in Section~\ref{sec:params}. Here, we present the other hyper-parameters of our method in Table~\ref{tab:params}, most of them follow Back-training~\cite{kulshreshtha2021back} and DPR~\cite{karpukhin2020dense}.

\section{Efficiency Report}

We list the time cost of training and inference in Table~\ref{tab:efficiency}. The evaluation is made with 8 NVIDIA A100 GPUs. The number of iterations is set as 3.

\section{Additional Experiments}

\begin{table}[t] \small
    \centering
    \caption{A filtered out generated query.}
    \begin{tabular}{p{0.9\columnwidth}}
        \toprule
        \textbf{Passage:} Baada ya uhuru wa Fiji (1970) kutoka kwa Waingereza, yalifuata mapinduzi ya kijeshi yaliyotokea mwaka 1987, hali iliyosababishwa na wakazi wa Fiji kulaumu serikali yao kutawaliwa na watu wa kabila la Indofijian au Wahindi. \\ \textbf{Translation:} After Fiji's independence (1970) from the British, it followed a military coup in 1987, a situation in which Fijians blamed their government for being ruled by Indofijian or Indian people. \\
        \midrule
        \textbf{Generated query:} Kwa upi uhuru wa Fiji ulifanyika mwaka gani? \\ \textbf{Translation:} In what year did Fiji's independence take place? \\
        \midrule
        \textbf{Top-1 passage:} Fiji ilijipatia uhuru wake kutoka katika utawala wa kikoloni wa Uingereza tarehe 10 Oktoba 1970. \\
        \textbf{Translation:} Fiji gained its independence from British colonial rule on October 10, 1970. \\
        \bottomrule
    \end{tabular}
    \label{tab:query-out}
\end{table}

\begin{table*}[t] \small
\centering
\caption{Effect of query filter. ``QF'' denotes query filter and ``HN'' denotes hard negative passages which are top passages returned by sparse and dense retrievers.}
\subtable[MRR@100]{\begin{tabular}{l|ccccccccccc|c} \toprule
Methods      & Ar             & Bn             & En             & Fi             & Id             & Ja             & Ko             & Ru             & Sw             & Te             & Th             & Avg           \\ \midrule
\name         & \textbf{58.5}  & \textbf{49.5}  & \textbf{37.3}  & \textbf{45.6}  & \textbf{51.3}  & \textbf{38.6}  & \textbf{43.6}  & \textbf{46.2}  & \textbf{48.6}  & \textbf{66.5}  & \textbf{53.1}  & \textbf{49.0} \\
w/o QF       & 53.5           & 47.1           & 33.1           & 39.9           & 47.5           & 36.6           & 36.2           & 44.0           & 44.2           & 27.6           & 42.0           & 41.1          \\
w/o QF \& HN & 58.1           & 48.6           & 37.0           & 45.2           & 50.6           & 38.4           & \textbf{43.6}  & 46.0           & 47.7           & 54.6           & 53.0           & 47.5          \\ \bottomrule
\end{tabular}}

\subtable[Recall@100]{\begin{tabular}{l|ccccccccccc|c} \toprule
Methods       & Ar             & Bn             & En             & Fi             & Id             & Ja             & Ko             & Ru             & Sw             & Te             & Th             & Avg           \\ \midrule
\name          & 85.9           & 89.2           & 77.6           & 83.1           & 86.4           & 77.0           & 74.9           & 81.8           & 83.4           & 95.6           & 85.3           & 83.6          \\
w/o QF        & 83.7           & 91.4           & 74.5           & 79.3           & 85.4           & 76.7           & 69.5           & 82.1           & 81.6           & 91.9           & 81.9           & 81.6          \\
w/o QF \& HN  & \textbf{86.2}  & \textbf{90.1}  & \textbf{79.1}  & \textbf{83.4}  & \textbf{87.3}  & \textbf{77.5}  & \textbf{76.8}  & \textbf{83.3}  & \textbf{82.9}  & \textbf{96.1}  & \textbf{86.7}  & \textbf{84.5} \\ \bottomrule
\end{tabular}}

\label{tab:query-lang}
\end{table*}

\subsection{Effect of Query Filter}
In our query generation module, we use the dense retriever to filter the generated queries. Here, we analyze the effectiveness of filtering based on mBERT. We present the result for {\ul w/o Queries Filter} and {\ul w/o Queries Filter \& Hard Negative passages} in Table~\ref{tab:query-lang}. As we can see, filtering generated queries can lead to better performance. We also find that w/o QF \& HN $>$ \name $>$ w/o QF on Recall@100. It denotes that the generated queries are relevant to the passages but the top passages returned by retrievers may be more relevant.
We present an example of a query that is filtered out in Table~\ref{tab:query-out}. 
As we can see, although the generated query can be answered by the passage, the main statement of the passage is not intended to answer the generated query and there is a more relevant passage to answer the generated query. As a result, these samples~(w/o hard negative passages) are helpful to Recall@100 but harmful to MRR@100.

\subsection{Visualization of the Training Procedure} 
\begin{figure}[t]
    \centering
    \includegraphics[width=0.85\columnwidth]{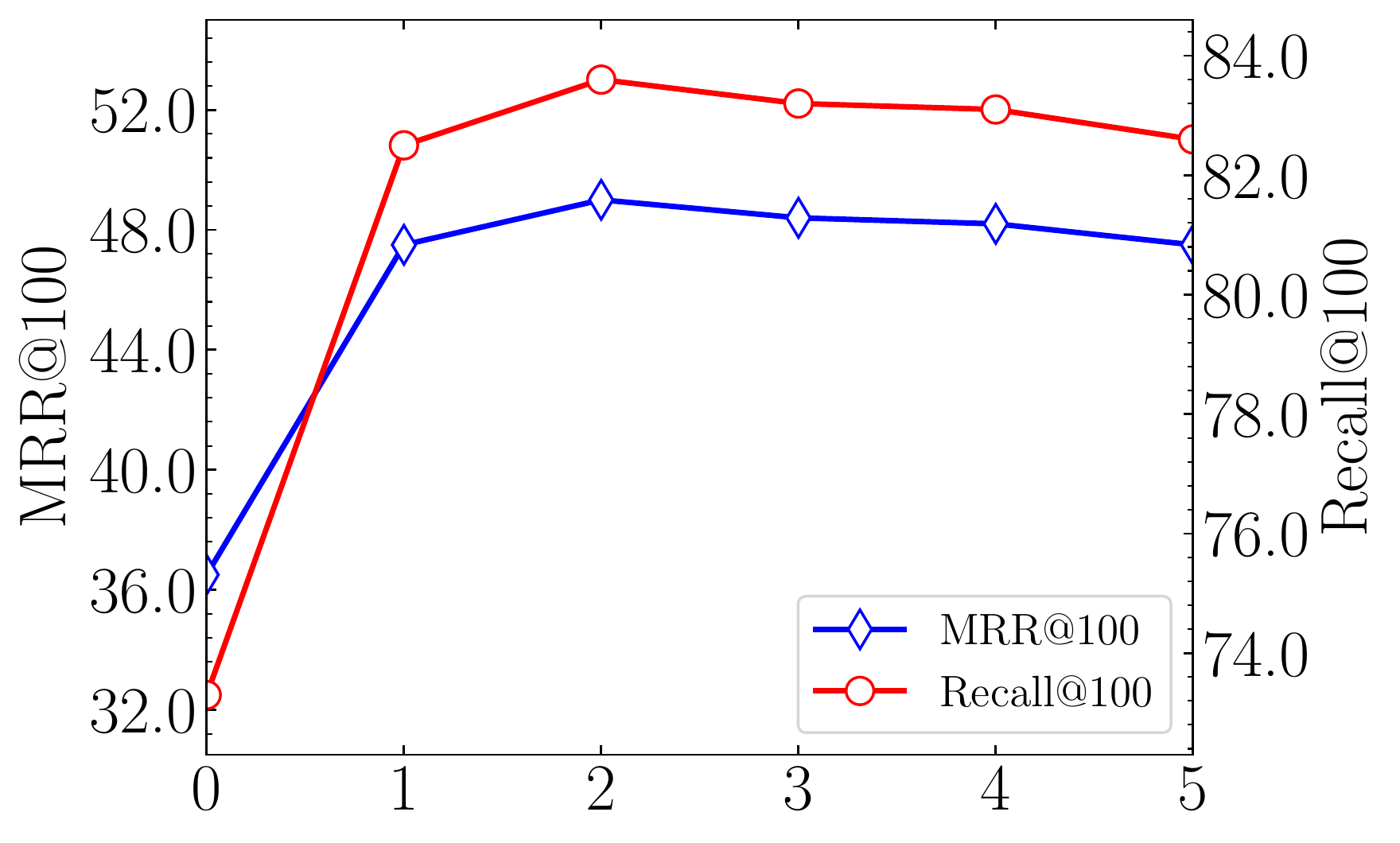}
    \caption{Iterative performance for the proposed \name based on mBERT.}
    \label{fig:iter}
\end{figure}

Our method employs iteratively training to improve the performance. Here, we report the iterative performance of our method in Figure~\ref{fig:iter}. To better show the effectiveness of our method, we set the number of iterations as 5 in the experiments. As we can see, the performance of our method increases with iteration and it holds steady when the model convergences. It shows that the distribution of mined data is similar to the distribution of real data, so the model does not suffer from the overfitting problem. In the end, the MRR@100 is improved by approximately 12\%, and Recall@100 is improved by approximately 10\%. It demonstrates the effectiveness of the mined data.

\subsection{Effect of English Data} \label{sec:nq}

In this section, we test the influence of different English data. As Xinyu \etal~\cite{zhang2022towards} said, MS-MARCO~\cite{bajaj2016ms} has a larger dataset size than NQ~\cite{kwiatkowski2019natural}, and the data size is a more critical factor. To test the effectiveness of our methods, we construct experiments on both MS-MARCO and NQ. Note that the only difference is the English data and we directly use the tuned parameters based on MS-MARCO for all experiments.
Table~\ref{tab:nq} presents the performance of two English data based on mBERT. Note that our re-implement mBERT based on NQ is better than the Mr. TYDI paper~\cite{zhang2021mr}\footnote{\href{https://github.com/castorini/mr.tydi}{https://github.com/castorini/mr.tydi}}. Because we share the parameters of the query encoder and the passage encoder, the trick leads to better performance.

As we can see, the gap between the performance of our method based on MS-MARCO data and NQ data is smaller than other methods. And the Recall@100 on NQ data is even higher than Recall@100 on MS-MACRO data. A possible reason is that NQ is closer to Mr. TYDI, both of them contain relatively well-formed queries posed against Wikipedia. The zero-shot performance of English data in Mr. TYDI data~(the Recall@100 of English is 75.1 for the model trained on MS-MARCO data and 78.3 for the model trained on NQ data) can demonstrate that. So, the mined data of dense retriever trained by NQ is more effective than MS-MARCO, and more effective data in target languages can lead to better performance.

\section{Detailed Results} \label{sec:lang}

Due to the limited space, we only present average performance in the experiment section. Here, we present the performance of each language. First, we present the detailed performance of both our method and baselines in Table~\ref{tab:mrtydi-lang}. Second, we present the detailed performance of ablation results in Table~\ref{tab:ablation-lang}. Finally, we present the detailed performance of variants of the passage mining module in Table~\ref{tab:passage-lang}. As we can see, our method performs better in most languages in these settings.

\begin{table*}[t] \footnotesize
\centering
\caption{Performance comparison for different English data based on mBERT.}

\subtable[MRR@100]{\begin{tabular}{c|ccccccccccc|c} \toprule
Methods       & Ar             & Bn             & En             & Fi             & Id             & Ja             & Ko             & Ru             & Sw             & Te             & Th             & Avg            \\ \midrule
\multicolumn{13}{c}{MS-MARCO} \\ \midrule
Zero-Shot     & 47.7           & 41.5           & 32.4           & 34.9           & 41.9           & 30.9           & 30.8           & 35.6           & 40.4           & 34.8           & 30.2           & 36.5           \\ \midrule 
Self-Training & 45.4           & 40.5           & 28.5           & 33.8           & 40.2           & 32.1           & 32.1           & 34.1           & 43.8           & 42.9           & 36.2           & 37.2           \\ 
Back-Training & 49.0           & 48.7           & 31.7           & 38.9           & 44.5           & 34.5           & 34.8           & 38.3           & 46.0           & 46.7           & 38.9           & 41.1           \\ 
\name          & \textbf{58.5}  & 49.5           & 37.3           & \textbf{45.6}  & \textbf{51.3}  & 38.6           & \textbf{43.6}  & \textbf{46.2}  & \textbf{48.6}  & 66.5           & \textbf{53.1}  & \textbf{49.0}  \\ \midrule
\multicolumn{13}{c}{NQ} \\ \midrule
Zero-Shot     & 30.0           & 38.7           & 30.6           & 25.6           & 30.7           & 30.6           & 23.4           & 29.6           & 28.1           & 24.1           & 22.6           & 28.5           \\ \midrule
Self-Training & 32.2           & 41.5           & 27.9           & 27.3           & 33.5           & 30.4           & 24.2           & 30.4           & 33.3           & 36.5           & 28.0           & 31.4           \\ 
Back-Training & 35.4           & 42.3           & 29.7           & 28.5           & 33.2           & 31.5           & 27.7           & 32.8           & 40.8           & 24.7           & 30.5           & 32.5           \\ 
\name          & 54.8           & \textbf{56.6}  & \textbf{41.0}  & 42.3           & 51.2           & \textbf{41.6}  & 38.9           & 46.1           & 46.2           & \textbf{66.7}  & 49.1           & 48.6           \\ \bottomrule
\end{tabular}}

\subtable[Recall@100]{\begin{tabular}{c|ccccccccccc|c} \toprule
Methods       & Ar             & Bn             & En             & Fi             & Id             & Ja             & Ko             & Ru             & Sw             & Te             & Th             & Avg            \\ \midrule
\multicolumn{13}{c}{MS-MARCO} \\ \midrule
Zero-Shot     & 80.6           & 78.8           & 75.1           & 74.7           & 79.3           & 67.8           & 65.5           & 73.1           & 70.4           & 77.1           & 63.7           & 73.3           \\ \midrule 
Self-Training & 82.3           & 86.9           & 74.4           & 78.2           & 81.3           & 72.7           & 67.0           & 77.6           & 78.7           & 91.0           & 73.6           & 78.5           \\ 
Back-Training & 84.6           & 90.1           & 76.5           & 81.4           & 84.4           & 76.2           & 73.6           & 82.2           & 80.8           & 89.6           & 83.1           & 82.0           \\ 
\name          & \textbf{85.9}  & 89.2           & 77.6           & 83.1           & 86.4           & 77.0           & 74.9           & 81.8           & 83.4           & 95.6           & 85.3           & 83.5           \\ \midrule
\multicolumn{13}{c}{NQ} \\ \midrule
Zero-Shot     & 70.3           & 80.2           & 78.3           & 69.2           & 76.4           & 74.3           & 60.9           & 72.7           & 65.7           & 69.9           & 59.9           & 70.7           \\ \midrule
Self-Training & 78.9           & 86.5           & 79.1           & 77.7           & 83.3           & 79.0           & 69.6           & 77.9           & 76.7           & 90.5           & 74.8           & 79.5           \\ 
Back-Training & 81.7           & 88.7           & 80.1           & 80.0           & 86.0           & 80.8           & 73.5           & 81.9           & 83.4           & 93.6           & 83.0           & 83.0           \\ 
\name          & \textbf{85.9}  & \textbf{91.9}  & \textbf{83.9}  & \textbf{84.3}  & \textbf{88.7}  & \textbf{83.2}  & \textbf{78.4}  & \textbf{85.5}  & \textbf{82.9}  & \textbf{95.7}  & \textbf{85.7}  & \textbf{86.0}  \\ \bottomrule
\end{tabular}}

\label{tab:nq}
\end{table*}

\begin{table*}[t] \small
\centering
\caption{Statistics for Mr. TYDI: number of queries (\# Q), judgments (\# J), and the number of passages.}
\begin{tabular}{l|rr|rr|rr|r} \toprule
    \multirow{2.5}{*}{L} & \multicolumn{2}{c|}{Train} & \multicolumn{2}{c|}{Dev} & \multicolumn{2}{c|}{Test} & \multirow{2.5}{*}{Corpus Size} \\ \cmidrule(lr){2-7}
    ~ & \# Q & \# J & \# Q & \# J & \# Q & \# J & ~ \\ \midrule
    Ar & 12,377 & 12,377 & 3,115 & 3,115 & 1,081 & 1,257 & 2,106,586 \\
    Bn & 1,713 & 1,719 & 440 & 443 & 111 & 130 & 304,059 \\
    En & 3,547 & 3,547 & 878 & 878 & 744 & 935 & 32,907,100 \\
    Fi & 6,561 & 6,561 & 1,738 & 1,738 & 1,254 & 1,451 & 1,908,757 \\
    Id & 4,902 & 4,902 & 1,224 & 1,224 & 829 & 961 & 1,469,399 \\
    Ja & 3,697 & 3,697 & 928 & 928 & 720 & 923 & 7,0000,027 \\
    Ko & 1,295 & 1,317 & 303 & 307 & 421 & 492 & 1,496,126 \\
    Ru & 5,366 & 5,366 & 1,375 &  1,375 & 955 & 1,168 & 9,597,504 \\
    Sw & 2,072 & 2,401 & 526 & 623 & 670 & 743 & 136,689 \\
    Te & 3,880 & 3,880 & 983 & 983 & 646 & 664 & 548,224 \\
    Th & 3,319 & 3,360 & 807 & 817 & 1,190 & 1,368 & 568,855 \\ \midrule
    All & 48,729 & 49,127 & 12,317 & 12,431 & 8,661 & 10,092 & 58,043,326 \\ \bottomrule
\end{tabular}
\label{tab:stat}
\end{table*}

\begin{table*}[t] \footnotesize
\centering
\caption{Detail results on Mr. TYDI test set.}

\subtable[MRR@100]{\begin{tabular}{c|c|ccccccccccc|c}

\toprule
\multicolumn{2}{c|}{Methods} 
                & Ar             & Bn             & En             & Fi             & Id             & Ja             & Ko             & Ru             & Sw             & Te             & Th             & Avg            \\
\midrule \multirow{2}{*}{\shortstack{Sparse \\ Retriever}}
& BM25          & 36.8           & 41.8           & 14.0           & 28.4           & 37.6           & 21.1           & 28.5           & 31.3           & 38.9           & 34.3           & 40.1           & 32.1           \\ 
& (tuned)       & 36.7           & 41.3           & 15.1           & 28.8           & 38.2           & 21.7           & 28.1           & 32.9           & 39.6           & 42.4           & 41.7           & 33.3           \\ \midrule 

\multirow{7}{*}{mBERT}
& Zero-Shot     & 44.4           & 38.3           & 31.5           & 30.6           & 37.8           & 31.4           & 29.7           & 33.7           & 36.9           & 36.3           & 28.2           & 34.4           \\  
& (reimpl)      & 47.7           & 41.5           & 32.4           & 34.9           & 41.9           & 30.9           & 30.8           & 35.6           & 40.4           & 34.8           & 30.2           & 36.5           \\ \cmidrule(lr){2-14}
& Self-Training & 45.4           & 40.5           & 28.5           & 33.8           & 40.2           & 32.1           & 32.1           & 34.1           & 43.8           & 42.9           & 36.2           & 37.2           \\ 
& Back-Training & 49.0           & 48.7           & 31.7           & 38.9           & 44.5           & 34.5           & 34.8           & 38.3           & 46.0           & 46.7           & 38.9           & 41.1           \\ 
& \name          & \textbf{58.5}  & \textbf{49.5}  & \textbf{37.3}  & \textbf{45.6}  & \textbf{51.3}  & \textbf{38.6}  & \textbf{43.6}  & \textbf{46.2}  & \textbf{48.6}  & \textbf{66.5}  & \textbf{53.1}  & \textbf{49.0}  \\ \cmidrule(lr){2-14}
& Supervised    & 64.2           & 55.2           & 45.4           & 52.7           & 53.8           & 43.1           & 42.8           & 46.4           & 58.8           & 83.3           & 54.5           & 54.6           \\ \midrule 

\multirow{6}{*}{XLM-R}
& Zero-Shot     & 31.7           & 40.9           & 18.1           & 30.3           & 31.7           & 22.9           & 31.7           & 27.0           & 24.4           & 36.0           & 39.5           & 30.4           \\ \cmidrule(lr){2-14}
& Self-Training & 37.8           & 41.6           & 21.6           & 31.6           & 36.0           & 25.6           & 29.4           & 28.1           & 40.5           & 54.6           & 38.4           & 35.0           \\  
& Back-Training & 32.0           & 43.0           & 20.2           & 28.7           & 31.3           & 25.1           & 30.4           & 29.0           & 20.2           & 27.5           & 38.8           & 29.6           \\ 
& \name          & \textbf{54.0}  & \textbf{51.7}  & \textbf{33.0}  & \textbf{43.9}  & \textbf{49.9}  & \textbf{34.8}  & \textbf{40.3}  & \textbf{41.4}  & \textbf{45.1}  & \textbf{70.1}  & \textbf{54.9}  & \textbf{47.2}  \\ \cmidrule(lr){2-14}
& Supervised    & 62.9           & 61.5           & 43.0           & 51.6           & 53.6           & 39.9           & 41.2           & 44.0           & 57.6           & 83.3           & 60.9           & 54.5           \\ \midrule 

\end{tabular}}

\subtable[Recall@100]{\begin{tabular}{c|c|ccccccccccc|c}

\toprule
\multicolumn{2}{c|}{Methods} 
                & Ar             & Bn             & En             & Fi             & Id             & Ja             & Ko             & Ru             & Sw             & Te             & Th             & Avg            \\
\midrule \multirow{2}{*}{\shortstack{Sparse \\ Retriever}}
& BM25          & 79.3           & 86.9           & 53.7           & 71.9           & 84.3           & 64.5           & 61.9           & 64.8           & 76.4           & 75.8           & 85.3           & 73.2           \\ 
& (tuned)       & 80.0           & 87.4           & 55.4           & 72.5           & 84.6           & 65.6           & 79.7           & 66.0           & 76.4           & 81.3           & 85.3           & 75.8           \\ \midrule 

\multirow{7}{*}{mBERT}
& Zero-Shot     & 79.9           & 82.0           & 75.8           & 69.3           & 75.8           & 73.8           & 64.5           & 72.8           & 68.6           & 79.7           & 64.8           & 73.4           \\  
& (reimpl)      & 80.6           & 78.8           & 75.1           & 74.7           & 79.3           & 67.8           & 65.5           & 73.1           & 70.4           & 77.1           & 63.7           & 73.3           \\ \cmidrule(lr){2-14}
& Self-Training & 82.3           & 86.9           & 74.4           & 78.2           & 81.3           & 72.7           & 67.0           & 77.6           & 78.7           & 91.0           & 73.6           & 78.5           \\ 
& Back-Training & 84.6           & \textbf{90.1}  & 76.5           & 81.4           & 84.4           & 76.2           & 73.6           & \textbf{82.2}  & 80.8           & 89.6           & 83.1           & 82.0           \\ 
& \name          & \textbf{85.9}  & 89.2           & \textbf{77.6}  & \textbf{83.1}  & \textbf{86.4}  & \textbf{77.0}  & \textbf{74.9}  & 81.8           & \textbf{83.4}  & \textbf{95.6}  & \textbf{85.3}  & \textbf{83.6}  \\ \cmidrule(lr){2-14}
& Supervised    & 90.2           & 92.3           & 84.2           & 85.8           & 87.7           & 82.1           & 78.8           & 84.7           & 85.9           & 96.2           & 88.7           & 87.0           \\ \midrule 

\multirow{6}{*}{XLM-R}
& Zero-Shot     & 76.3           & 84.2           & 68.8           & 74.6           & 82.3           & 66.7           & 67.9           & 68.9           & 60.3           & 81.2           & 86.5           & 74.3           \\ \cmidrule(lr){2-14}
& Self-Training & 78.9           & 85.1           & 68.8           & 78.6           & 84.0           & 70.2           & 66.8           & 71.6           & 78.2           & 93.6           & 88.4           & 78.6           \\  
& Back-Training & 78.9           & \textbf{88.3}  & 72.4           & 77.3           & 84.5           & 73.1           & 71.1           & 75.2           & 59.7           & 80.6           & 91.1           & 77.5           \\ 
& \name          & \textbf{85.0}  & 86.0           & \textbf{76.2}  & \textbf{82.6}  & \textbf{86.4}  & \textbf{74.6}  & \textbf{73.6}  & \textbf{77.8}  & \textbf{80.2}  & \textbf{95.4}  & \textbf{92.1}  & \textbf{82.7} \\ \cmidrule(lr){2-14}
& Supervised    & 89.3           & 92.8           & 83.1           & 86.3           & 89.8           & 80.1           & 78.7           & 82.6           & 87.0           & 96.7           & 92.6           & 87.2           \\ \midrule 

\end{tabular}}

\label{tab:mrtydi-lang}
\end{table*}

\begin{table*}[t] \small
\centering
\caption{Ablation results based on mBERT. }
\subtable[MRR@100]{\begin{tabular}{l|ccccccccccc|c} \toprule
Methods      & Ar             & Bn             & En             & Fi             & Id             & Ja             & Ko             & Ru             & Sw             & Te             & Th             & Avg           \\ \midrule
All          & 58.5           & 49.5           & \textbf{37.3}  & \textbf{45.6}  & \textbf{51.3}  & \textbf{38.6}  & \textbf{43.6}  & \textbf{46.2}  & \textbf{48.6}  & 66.5           & \textbf{53.1}  & \textbf{49.0} \\ \midrule
w/o LR       & 55.7           & \textbf{49.6}  & 35.9           & 43.0           & 49.7           & 38.1           & 40.1           & 43.5           & 46.9           & \textbf{66.9}  & 45.9           & 46.9          \\
w/o LR$_+$   & 49.1           & 45.5           & 32.4           & 36.4           & 43.4           & 32.0           & 32.4           & 39.0           & 41.9           & 45.9           & 39.0           & 39.7          \\
w/o QG       & \textbf{58.8}  & 49.4           & 37.0           & 45.1           & 51.0           & 38.0           & 42.7           & 45.5           & 47.2           & 62.7           & 51.4           & 48.1          \\
w/o LR + QG  & 47.7           & 41.5           & 32.4           & 34.9           & 41.9           & 30.9           & 30.8           & 35.6           & 40.4           & 34.8           & 30.2           & 36.5          \\ \bottomrule
\end{tabular}}

\subtable[Recall@100]{\begin{tabular}{l|ccccccccccc|c} \toprule
Methods      & Ar             & Bn             & En             & Fi             & Id             & Ja             & Ko             & Ru             & Sw             & Te             & Th             & Avg           \\ \midrule
All          & \textbf{85.9}  & \textbf{89.2}  & 77.6           & 83.1           & 86.4           & \textbf{77.0}  & \textbf{74.9}  & 81.8           & \textbf{83.4}  & \textbf{95.6}  & \textbf{85.3}  & \textbf{83.6} \\ \midrule
w/o LR       & 84.0           & 87.4           & 76.0           & 80.4           & 85.1           & 76.2           & 70.7           & 80.7           & 81.7           & 93.8           & 81.6           & 81.6          \\
w/o LR$_+$   & 80.8           & 85.1           & 73.7           & 76.8           & 82.5           & 71.1           & 65.7           & 77.7           & 77.5           & 89.5           & 76.6           & 77.9          \\
w/o QG       & 85.7           & 88.3           & \textbf{78.6}  & \textbf{83.3}  & \textbf{86.9}  & 76.2           & 74.4           & 81.7           & 80.8           & 95.4           & 84.3           & 83.2          \\
w/o LR + QG  & 80.6           & 78.8           & 75.1           & 74.7           & 79.3           & 67.8           & 65.5           & 73.1           & 70.4           & 77.1           & 63.7           & 73.3          \\ \bottomrule
\end{tabular}}

\label{tab:ablation-lang}
\end{table*}

\begin{table*}[t] \small
\centering
\caption{Effect of lexicon-enhanced retrieval module.}
\subtable[MRR@100]{\begin{tabular}{l|ccccccccccc|c} \toprule
Methods                 & Ar             & Bn             & En             & Fi             & Id             & Ja             & Ko             & Ru             & Sw             & Te             & Th             & Avg  \\ \midrule
Sparse                  & 36.8           & 41.8           & 14.0           & 28.4           & 37.6           & 21.1           & 28.5           & 31.3           & 38.9           & 34.3           & 40.1           & 32.1 \\ 
Dense                   & 47.7           & 41.5           & 32.4           & 34.9           & 41.9           & 30.9           & 30.8           & 35.6           & 40.4           & 34.8           & 30.2           & 36.5 \\ \midrule
\name w/o QG             & \textbf{58.5}  & 50.8           & \textbf{37.0}  & \textbf{45.0}  & \textbf{52.1}  & \textbf{38.2}  & \textbf{41.7}  & \textbf{45.0}  & 46.6           & \textbf{59.9}  & 48.2           & \textbf{47.5}  \\ \midrule
Sparse $+$ Dense        & 57.1           & \textbf{54.1}  & 34.3           & 43.6           & 50.2           & 37.4           & 38.6           & 44.3           & \textbf{47.7}  & 55.7           & \textbf{49.9}  & 46.6  \\
Sparse $\times$ Dense   & 48.5           & 48.3           & 25.7           & 37.1           & 44.0           & 31.2           & 34.6           & 39.0           & 42.4           & 49.8           & 47.8           & 40.8  \\
Double Dense Retrievers               & 52.5           & 52.5           & 36.4           & 40.6           & 48.2           & 37.9           & 38.1           & 40.8           & 45.1           & 56.4           & 40.7           & 44.5  \\ \midrule
w/o HN                  & 52.2           & 47.4           & 32.6           & 39.9           & 46.7           & 35.3           & 36.6           & 39.8           & 45.5           & 53.2           & 38.5           & 42.5  \\ 
w/ Sparse HN            & 54.3           & 45.6           & 34.8           & 41.5           & 49.3           & 35.7           & 37.1           & 41.5           & 46.1           & 50.9           & 39.9           & 43.3  \\ \bottomrule
\end{tabular}}

\subtable[Recall@100]{\begin{tabular}{l|ccccccccccc|c} \toprule
Methods                 & Ar             & Bn             & En             & Fi             & Id             & Ja             & Ko             & Ru             & Sw             & Te             & Th             & Avg  \\ \midrule
Sparse                  & 79.3           & 86.9           & 53.7           & 71.9           & 84.3           & 64.5           & 61.9           & 64.8           & 76.4           & 75.8           & 85.3           & 73.2 \\
Dense                   & 80.6           & 78.8           & 75.1           & 74.7           & 79.3           & 67.8           & 65.5           & 73.1           & 70.4           & 77.1           & 63.7           & 73.3 \\ \midrule
\name w/o QG             & \textbf{85.9}  & 89.2           & 78.9           & \textbf{82.5}  & \textbf{85.8}  & 75.2           & 73.6           & \textbf{81.3}  & \textbf{79.3}  & \textbf{94.0}  & \textbf{81.9}  & \textbf{82.5}  \\ \midrule
Sparse $+$ Dense        & 85.1           & \textbf{89.6}  & 77.0           & 79.8           & \textbf{85.8}  & \textbf{77.4}  & \textbf{75.5}  & 79.3           & 78.3           & 85.8           & 78.7           & 81.1  \\
Sparse $\times$ Dense   & 84.5           & \textbf{89.6}  & 75.3           & 78.9           & 85.0           & 76.1           & 75.1           & 78.6           & 76.5           & 85.7           & 78.7           & 80.4  \\
Double Dense Retrievers             & 84.5           & 88.7           & \textbf{79.5}  & 81.7           & 85.5           & 77.1           & 73.9           & 79.7           & \textbf{79.3}  & 92.2           & 78.2           & 81.9  \\ \midrule
w/o HN                  & 84.8           & 86.9           & 77.0           & 80.6           & 83.0           & 74.2           & 72.4           & 80.2           & 77.5           & 92.2           & 75.6           & 80.4  \\ 
w/ Sparse HN            & 84.2           & 83.3           & 77.4           & 80.2           & 83.8           & 71.8           & 70.3           & 78.1           & 77.4           & 89.9           & 74.7           & 79.2  \\ \bottomrule
\end{tabular}}
\label{tab:passage-lang}
\end{table*}

\end{document}